\documentclass[aps, showpacs, showkeys, nofootinbib, floatfix]{revtex4}
\usepackage{amssymb}
\usepackage{amsmath}
\usepackage{graphicx}
\usepackage{subfigure} 

\begin{document}

\title{Cosmological constraints on coupled dark energy}

\author{Weiqiang Yang$^a$\footnote{d11102004@mail.dlut.edu.cn},
Hang Li$^b$\footnote{lihang@semi.ac.cn},
Yabo Wu$^a$, Jianbo Lu$^a$}

\affiliation{$^a$Department of Physics, Liaoning Normal University, Dalian, 116029, P. R. China \\
$^b$College of Medical Laboratory, Dalian Medical University, Dalian, 116044, P. R. China
}

\begin{abstract}
The coupled dark energy model provides a possible approach to mitigate the coincidence problem of cosmological standard model. Here, the coupling term is assumed as $\bar{Q}=3H\xi_x\bar{\rho}_x$, which is related to the interaction rate and energy density of dark energy. We derive the background and perturbation evolution equations for several coupled models. Then, we test these models by currently available cosmic observations which include cosmic microwave background radiation from Planck 2015, baryon acoustic oscillation, type Ia supernovae, $f\sigma_8(z)$ data points from redshift-space distortions, and weak gravitational lensing. The constraint results tell us there is no evidence of interaction at 2$\sigma$ level, it is very hard to distinguish different coupled models from other ones.
\end{abstract}

\maketitle

\section{Introduction}

After the Planck 2015 data sets \cite{ref:Planck2015-1,ref:Planck2015-2} have been released, the cosmological constraints on $\Lambda$CDM model \cite{ref:Planck2015-3} tell us that dark energy occupies about $68.5\%$ of the Universe, dark matter accounts for $27.1\%$, and baryonic matter occupies $4.4\%$. The $\Lambda$CDM model is in good agreement with the cosmic observational data, however, it encounters the coincidence problem \cite{ref:Zlatev1999}. An effective method to alleviate this issue is considering the interaction between dark matter and dark energy. The coupling of dark sectors would influence the background evolution of the Universe and affect the growth history of the cosmic structure. Up to now, it is difficult to identify the coupled form from the fundamental theory. Thus, the interacting dark energy models are mostly based on the phenomenological consideration. A kind of popular interaction forms is related to the energy densities of the dark components \cite{ref:Potter2011,ref:Aviles2011,ref:Caldera-Cabral2009,ref:Boehmer2010,ref:Song2009,ref:Koyama2009,ref:Majerotto2010,
ref:Valiviita2010,ref:Valiviita2008,ref:Jackson2009,ref:Clemson2012,ref:Bean2008,ref:Gavela2009,ref:Gavela2010,ref:Yang2014-uc,ref:Yang2014-ux,ref:Yang2014-dh,
ref:Salvatelli2013,ref:Quartin2008,ref:Honorez2010,ref:Bernardis2011,ref:He2010,ref:Abdalla2009,ref:Sadjadi2010,ref:Olivares2008,ref:Sun2013,ref:Sadjadi2006,
ref:Sadeghi2013,ref:Zhang2013,ref:Koivisto2005,ref:Simpson2011,ref:Bertolami2007,ref:Avelino2012,ref:Quercellini2008,ref:Mohammadi2012,ref:Sharif2012,ref:Wu2007,
ref:Barrow2006,ref:Zimdahl2001,ref:Lip2011d,ref:Chen2009,ref:Koshelev2009,ref:Zhang2012,ref:Cao2011,ref:Guo2007,ref:Liyh2013,ref:Bolotin2013,ref:Chimento2013,
ref:Forte2013,ref:Baldi2012,ref:Xia2013,ref:Amendola2006,
ref:Ng2006,ref:Ng2010,ref:Kumar2016,ref:Bamba2012,ref:Amico2016,ref:Marsh2016,ref:Boriero2015,
ref:Valiviita2015,ref:Murgia2016,ref:Martin2015,ref:Fay2016,ref:Goncalves2015,ref:Pourtsidou2016,ref:Skordis2015,ref:Yang-Gauss2015,ref:Duniya2015,
ref:Wang2015,ref:Liyh2016,ref:Marcondes2016,ref:Ferreira2014,ref:Costa2016,ref:WangBin2016,ref:wyt2015,ref:Nunes2016}.

The coupling between dark sectors could significantly affect the growth history of cosmic structure, one can see Refs. \cite{ref:Clemson2012,ref:Caldera-Cabral2009,ref:Song2009,ref:Koyama2009,ref:Honorez2010,ref:Koshelev2009}. In the test of galaxy clustering, the redshifts need to be translated to distances, so the measured clustering would be highly anisotropic. An important source of this anisotropy are redshift-space distortions (RSD) \cite{ref:RSD-Kaiser1987}. RSD arise because peculiar velocities contribute to observed galaxies redshifts, a spherical overdensity appears distorted by peculiar velocities when observed in the redshift space. On linear scales, the overdenstiy appears squashed along the line of sight. For a detailed review of RSD, one can see Ref. \cite{ref:RSD-Hamilton1998}. RSD allow measurements of the amplitude of fluctuations in the velocity field, in linear theory, a model-dependent measurement of $f\sigma_8(z)$ has been suggested in Ref. \cite{ref:fsigma8-DE-Song2009}. The measures of the distribution of matter could be the distribution of galaxy distribution and statistics of weak lensing. The galaxy distribution could be measured by RSD. However, the galaxy distribution has crucial disadvantage due to the bias problem, $\delta_g=b\delta_{c}$, i.e. the density contrast of galaxy is proportional to that of cold dark matter. On large scales where the density contrast is small, the relation is viable, but on small scales the relation is not valid and the bias largely depends on the survey thus on the galaxy population and the scale $k$. Therefore the galaxy distribution is not reliable or robust tracer for the matter distribution. Whereas the convergence of weak gravitational lensing never suffers from the bias problem. Weak gravitational lensing of the images of distant galaxies offers a useful way to map the matter distribution in the Universe. The cosmological information can be extracted through the two-point shear correlation function which is related to convergence power spectrum via $\xi^{i,j}_{\pm}(\theta)=\frac{1}{2\pi}\int^\infty_0 P^{i,j}_{\kappa}(l)J_{\pm}(l\theta)ldl$ \cite{ref:Bartelmann2011}, where $J_{\pm}(l\theta)$ is the zeroth (for $\xi_+$) and fourth (for $\xi_-$) order Bessel function of the first kind. $P_{\kappa}(l)$ is the convergence power spectrum at angular wave number $l$. Ref. \cite{ref:Heymans2013} presents a finely-binned tomographic weak lensing (WL) analysis of the Canada-France-Hawaii Telescope Lensing Survey (CFHTLenS), mitigating contamination to the signal from the presence of intrinsic galaxy alignments via the simultaneous fit of a cosmological model and an intrinsic alignment model. Here, in order to obtain reliable constraint results of parameters space, we would test the coupled dark energy models from both RSD and WL data sets on large scale structure. It is worthwhile to anticipate that the large scale structure measurement will help to significantly tighten the cosmological constraints and some evidences could rule out large interaction rate.

In this paper, the coupling term is assumed as $\bar{Q}=3H\xi_x\bar{\rho}_x$, which is free from the large scale instability of the perturbation \cite{ref:Valiviita2008}. For the Hubble expansion rate, $H$ denotes the total expansion rate (background plus perturbations), $H=\bar{H}+\delta H$ \cite{ref:Gavela2010}, which could influence the perturbation part of the interaction term, furthermore it would impact the continuity and Euler equations of interacting dark sectors. For the momentum transfer, the simplest physical choice is that there is no momentum transfer in the rest frame of either dark matter or dark energy \cite{ref:Clemson2012}. It is noteworthy that the Euler equation of dark matter would be modified in the rest frame of dark energy, the dark matter no longer follows geodesics in general \cite{ref:Koyama2009}. We need pay attention to the scale-dependence of the density perturbations of dark matter, Ref. \cite{ref:Yang2014-ux} have analysed this issue and concluded that the growth rate of matter is scale-independent on the linear scale, we could safely test the interaction rate with the $f\sigma_8(z)$ data sets. Moreover, one could judge the stability of the perturbations via the doom factor of coupled model \cite{ref:Gavela2009}. The doom factor is defined as $d\equiv\frac{-\bar{Q}}{3H\bar{\rho}_x(1+w_x)}=\frac{-\xi_x}{1+w_x}$, where $w_x$ is the equation of state parameter of dark energy. According to the conclusion of Refs. \cite{ref:Gavela2009,ref:Clemson2012}: when $d<0$, the stable perturbations could be acquired for the interacting form $\bar{Q}=3H\xi_x\bar{\rho}_x$. It means that the perturbation stability requires the conditions $\xi_x>0$ and $(1+w_x)>0$ or $\xi_x<0$ and $(1+w_x)<0$. Here, in order to avoid the phantom doomsday \cite{ref:Caldwell2003}, we would discuss the stable case of $\xi_x>0$ and $(1+w_x)>0$.

This paper is organized as follows. In Sec. II, in the rest frame of dark matter or dark energy, we introduce several interacting dark energy models with or without perturbed Hubble expansion rate. We present the background behavior and the first-order perturbation equations in different cases. In Sec. III, based on the large scale structure measurements (RSD and WL), we test the parameters space of interacting dark energy models. The last section is the conclusion of this paper.

\section{The background and perturbation equations of coupled models}

In a flat Friedman-Robertson-Walker (FRW) universe, according to the phenomenological approach, the coupling could be introduced into the background conservation equations of dark matter and dark energy
\begin{eqnarray}
\bar{\rho}'_c+3\mathcal{H}\bar{\rho}_c=a\bar{Q}_c=-a\bar{Q}, \\
\bar{\rho}'_x+3\mathcal{H}(1+w_x)\bar{\rho}_x=a\bar{Q}_x=a\bar{Q},
\end{eqnarray}
where the subscript $c$ and $x$ respectively stand for dark matter and dark energy, the prime denotes the derivative with respect to conformal time $\tau$, $a$ is the scale factor of the Universe, $\mathcal{H}=a'/a=a\bar{H}$ is the conformal Hubble parameter, $\bar{Q}>0$ presents that the direction of energy transfer is from dark matter to dark energy; $\bar{Q}<0$ means the opposite direction of the energy exchange.

In a general gauge, the perturbed FRW metric is \cite{ref:Majerotto2010,ref:Valiviita2008,ref:Clemson2012}
\begin{eqnarray}
ds^2=a^2(\tau)\{ -(1+2\phi)d\tau^2+2\partial_iBd\tau dx^i+[(1-2\psi)\delta_{ij}+2\partial_i\partial_jE]dx^idx^j \},
\label{eq:per-metric}
\end{eqnarray}
where $\phi$, B, $\psi$ and E are the gauge-dependent scalar perturbations quantities.

The general forms for density perturbations (continuity) and velocity perturbations (Euler) equations of A fluid \cite{ref:Majerotto2010,ref:Valiviita2008,ref:Clemson2012,ref:Xu2011}
\begin{eqnarray}
\delta'_A+3\mathcal{H}(c^2_{sA}-w_A)\delta_A
+9\mathcal{H}^2(1+w_A)(c^2_{sA}-c^2_{aA})\frac{\theta_A}{k^2}
+(1+w_A)\theta_A-3(1+w_A)\psi'+(1+w_A)k^2(B-E')
\nonumber \\
=\frac{a}{\bar{\rho}_A}(-\bar{Q}_A\delta_A+\delta Q_A)
+\frac{a\bar{Q}_A}{\bar{\rho}_A}\left[\phi+3\mathcal{H}(c^2_{sA}-c^2_{aA})\frac{\theta_A}{k^2}\right],
\label{eq:general-deltaA}
\end{eqnarray}
\begin{eqnarray}
\theta'_A+\mathcal{H}(1-3c^2_{sA})\theta_A-\frac{c^2_{sA}}{(1+w_A)}k^2\delta_A
-k^2\phi
=\frac{a}{(1+w_A)\bar{\rho}_A}[(\bar{Q}_A\theta-k^2f_A)-(1+c^2_{sA})\bar{Q}_A\theta_A],
\label{eq:general-thetaA}
\end{eqnarray}
where $\delta_A=\delta\rho_A/\bar{\rho}_A$ is the density contrast of A fluid, $\theta_A=-k^2(v_A+B)$ is the volume expansion of A fluid in Fourier space \cite{ref:Valiviita2008,ref:Ma1995}, $\theta$ is the volume expansion of total fluid, $v_A$ is the peculiar velocity potential, $k$ is the wavenumber; $c^2_{aA}$ is the adiabatic sound speed whose definition is $c^2_{aA}=\bar{p}'_A/\bar{\rho}'_A=w_x+w'_x/(\bar{\rho}'_A/\bar{\rho}_A)$, and $c^2_{sA}$ is the physical sound speed in the rest frame, its definition is $c^2_{sA}=(\delta p_A/\delta\rho_A)_{restframe}$ \cite{ref:Valiviita2008,ref:Kodama1984,ref:Hu1998,ref:Gordon2004}. In order to avoid the unphysical instability, $c^2_{sA}$ should be taken as a non-negative parameter \cite{ref:Valiviita2008}.

We introduce a simple parameter $b_1$ to "choose the rest frame of dark matter or dark energy",
\[b_1 = \left\{ \begin{array}{l}
 1,~~for~~Q^{\mu}~\parallel~u^{\mu}_{(c)}, \\
 0,~~for~~Q^{\mu}~\parallel~u^{\mu}_{(x)}, \\
 \end{array} \right.\]
where $Q^{\mu}$ is the energy transfer four-vector, $u^{\mu}_{(i)}(i=c,x)$ is the four-velocity of dark matter or dark energy. After determining the parameter $b_1$, the momentum transfer potential $f_A$ could be assumed that $k^2f_A=\bar{Q}_A[\theta-b\theta_c-(1-b)\theta_x]$ in the rest frame of dark matter, where the energy-momentum transfer four-vector $Q^{\mu}_A$ is relative to the four-velocity $u^{\mu}$, and it can be split as $Q^A_0=-a[\bar{Q}_A(1+\phi)+\delta Q_A]$, $Q^A_i=a\partial_i[\bar{Q}_A(v+B)+f_A]$ \cite{ref:Majerotto2010,ref:Valiviita2008,ref:Clemson2012}. Besides, when the perturbed Hubble expansion rate is considered in the perturbation equations of dark sectors, $H$ denotes the total expansion rate (background plus perturbations), $H=\bar{H}+\delta H$, we also introduce another simple parameter $b_2$ to determine the coupled models with perturbed $H$ ($b_2=1$) or without perturbed $H$ ($b_2=0$). According to the analysis on the contribution from the expansion rate perturbation $\delta H/\bar{H}$ in Ref. \cite{ref:Gavela2010}, $\delta H/\bar{H}=(\theta+h'/2)/(3\mathcal{H})$. Moreover, in light of $(\rho+p)v=\sum (\rho_A+p_A)v_A$ \cite{ref:Valiviita2008,ref:Gavela2010}, we would obtain the continuity and Euler equations of different interaction cases

\begin{eqnarray}
\delta'_x+(1+w_x)\left(\theta_x+\frac{h'}{2}\right)+3\mathcal{H}(c^2_{sx}-w_x)\delta_x
+9\mathcal{H}^2(c^2_{sx}-w_x)(1+w_x)\frac{\theta_x}{k^2}
\nonumber \\
=9\mathcal{H}^2(c^2_{sx}-w_x)\xi_x\frac{\theta_x}{k^2}
+b_2\xi_x\left(\theta+\frac{h'}{2}\right),
\label{eq:per-deltax}
\end{eqnarray}
\begin{eqnarray}
\delta'_c+\theta_c+\frac{h'}{2}
=3\mathcal{H}\xi_x\frac{\rho_x}{\rho_c}(\delta_c-\delta_x)
-b_2\xi_x\frac{\rho_x}{\rho_c}\left(\theta+\frac{h'}{2}\right),
\label{eq:per-deltac}
\end{eqnarray}
\begin{eqnarray}
\theta'_x+\mathcal{H}(1-3c^2_{sx})\theta_x-\frac{c^2_{sx}}{1+w_x}k^2\delta_x
=\frac{3\mathcal{H}\xi_x}{1+w_x}
[b_1(\theta_c-\theta_x)-c^2_{sx}\theta_x],
\label{eq:per-thetax}
\end{eqnarray}
\begin{eqnarray}
\theta'_c+\mathcal{H}\theta_c
=3\mathcal{H}\xi_x\frac{\rho_x}{\rho_c}(1-b_1)(\theta_c-\theta_x).
\label{eq:per-thetac}
\end{eqnarray}

The above perturbation equations include four interacting dark energy models, the first or second one is the coupled model with perturbed expansion rate $(b_2=1)$ or without perturbed expansion rate $(b_2=0)$ in the rest frame of dark matter ($b_1=1$ for $Q^{\mu}\parallel u^{\mu}_{(c)}$), the third or fourth one is the coupled model with perturbed expansion rate $(b_2=1)$ or without perturbed expansion rate $(b_2=0)$ in the rest frame of dark energy ($b_1=0$ for $Q^{\mu}\parallel u^{\mu}_{(x)}$). According to this order, the four interacting dark energy models are named as IDE1 (the first interacting dark energy model), IDE2, IDE3, and IDE4 model. In the next section, we would pay attention to the constraint results of the model parameter space. Moreover, we try to find the difference from the four interaction cases.

\section{Observational data sets and cosmological constraint results}

For the coupled models, we consider the following eight-dimensional parameter space
\begin{eqnarray}
P\equiv\{\Omega_bh^2, \Omega_{c}h^2, \Theta_S, \tau, w_x, \xi_x, n_s, log[10^{10}A_S]\},
\label{eq:parameter_space}
\end{eqnarray}
the priors of the basic model parameters are shown in the second column of Table \ref{tab:results-mean}. The pivot scale of the initial scalar power spectrum $k_{s0}=0.05Mpc^{-1}$ is adopted. Moreover, the priors of the cosmic age $10Gyr<t_0<20Gyr$ and Hubble constant $H_0=73.8\pm2.4 km s^{-1} Mpc^{-1}$ \cite{ref:Riess2011} are used. In order to avoid the unphysical sound speed, we assume $c^2_{sx}=1$ according to Refs. \cite{ref:Valiviita2008,ref:Majerotto2010,ref:Clemson2012}.

For our numerical calculations, the total likelihood $\chi^2$ can be constructed as
\begin{eqnarray}
\chi^2=\chi^2_{CMB}+\chi^2_{BAO}+\chi^2_{SNIa}+\chi^2_{RSD}+\chi^2_{WL},
\label{eq:chi2}
\end{eqnarray}
where the four terms in right side of this equation, respectively, denote the contribution from CMB, BAO (baryon acoustic oscillations), SNIa (type-Ia supernovae), RSD, and WL data sets. The used data sets for our Monte Carlo Markov Chain (MCMC) likelihood analysis are listed in Table \ref{tab:alldata}. We modified the public available CosmoMC packages \cite{ref:cosmomc-Lewis2002} in order to test the model parameter space.

\begin{table}
\begin{center}
\begin{tabular}{ccc}
\hline\hline Data names & Data descriptions and references \\ \hline
CMB & full Planck temperature-only $C^{TT}_l$ and the low$-l$ polarization $C^{TE}_l+C^{EE}_l+C^{BB}_l$ \cite{ref:Planck2015-2} \\
BAO & $r_s/D_V(z=0.106, 0.35, 0.57)=0.336\pm0.015, 0.1126\pm0.0022, 0.0732\pm0.0012$\cite{ref:BAO-1,ref:BAO-2,ref:BAO-3} \\
SNIa & the Joint Light-curve Analysis (JLA) sample \cite{ref:JLA} \\
RSD & $f\sigma_8(z)$ data points from redshift-space distortions \cite{ref:fsigma83-Samushia2012} \\
WL & the blue galaxy sample on the WL analysis of CFHTLenS \cite{ref:Heymans2013,ref:Heymans2016} \\
\hline\hline
\end{tabular}
\caption{The used data sets for our MCMC likelihood analysis on the interacting dark energy models.}
\label{tab:alldata}
\end{center}
\end{table}

\begin{figure}
\centering
\subfigure{
\begin{minipage}[b]{0.48\textwidth}
\includegraphics[width=1\textwidth]{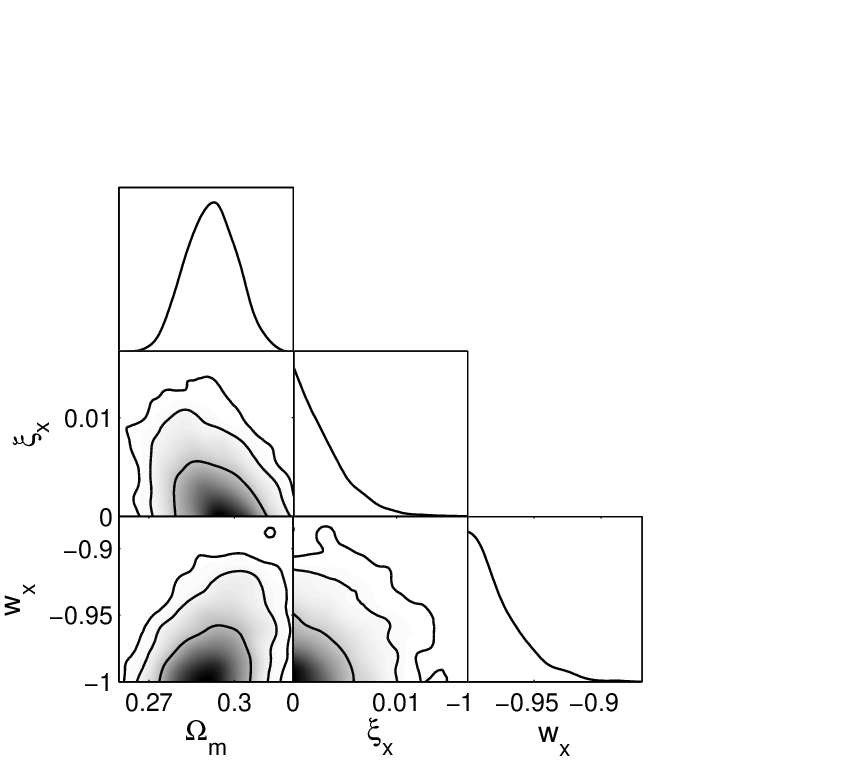} \\ 
\includegraphics[width=1\textwidth]{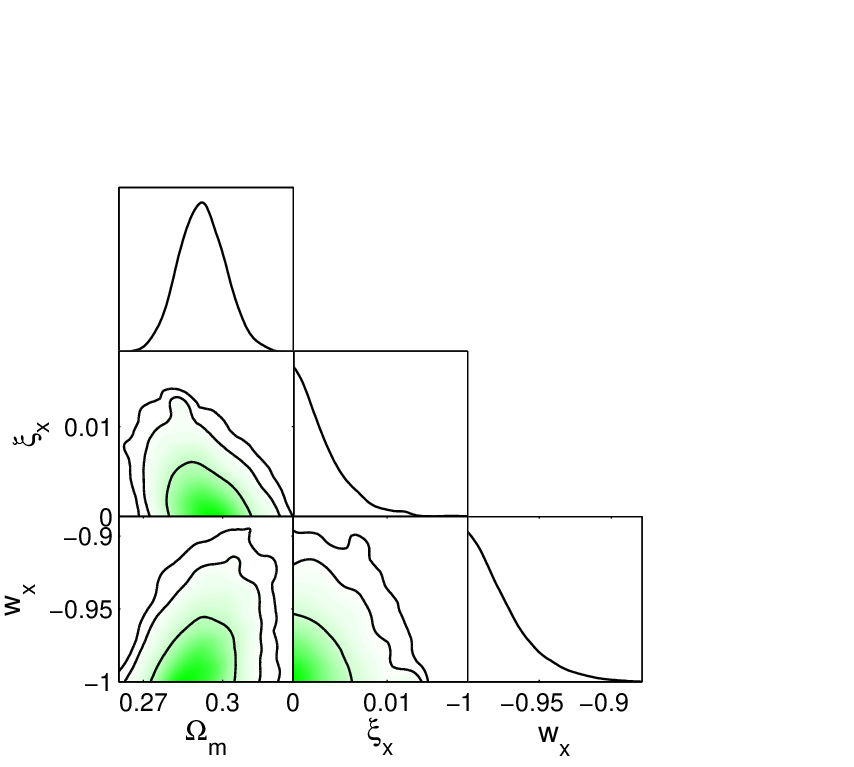}    
\end{minipage}
}
\subfigure{
\begin{minipage}[b]{0.48\textwidth}
\includegraphics[width=1\textwidth]{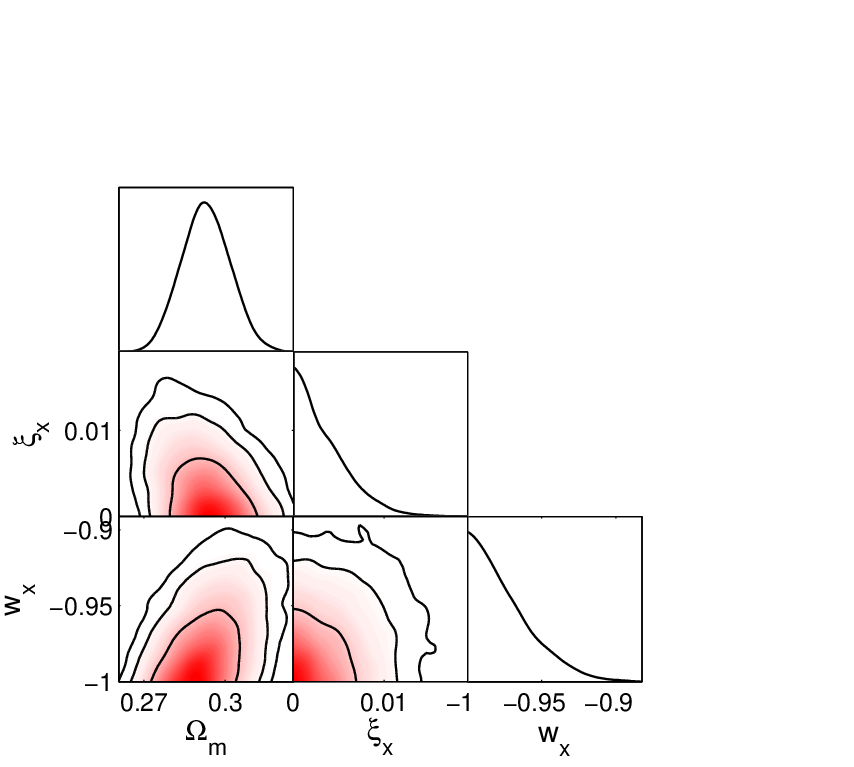} \\ 
\includegraphics[width=1\textwidth]{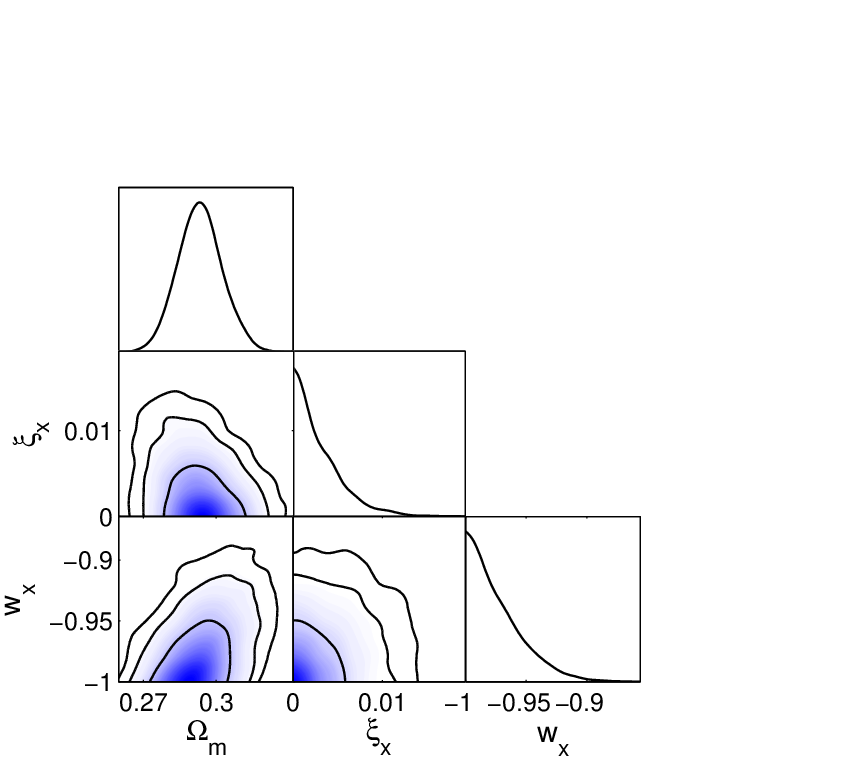}    
\end{minipage}
}
\caption{The 1D marginalized distributions on the parameters $\xi_x$, $w_x$, $\Omega_m$ and 2D contours of the four coupled models with 68\% C.L., 95 \% C.L., and 99.7\% C.L. between each other, where CMB from Planck 2015, BAO, SNIa, RSD, and WL data sets have been used. The color map of IDE1, IDE2, IDE3, IDE4 are, respectively, black, red, green, blue.}
\label{fig:contour}
\end{figure}

After running eight chains in parallel, in order to clearly see that the difference from the four coupled models, we list the mean values of the basic and derived parameters for four coupled models with $1,2,3\sigma$ errors in Table \ref{tab:results-mean}, the constraint results are almostly the same at $2\sigma$ level for the four interaction cases. In Figs. \ref{fig:contour}, We present the one-dimensional (1D) marginalized distributions of the parameters $\xi_x$, $w_x$, $\Omega_m$ and two-dimensional (2D) contours with $68\%$ confidence level (C.L.), $95\%$ C.L., and $99.7\%$ C.L. From these figures, comparing the results obtained in the coupled model with other ones, it is also very hard to find some difference from several interacting dark energy models. Especially, we pay attention to the interaction rate in the coupled dark energy model. Using CMB from Planck 2015, BAO, SNIa, RSD, WL measurements, the results showed the interaction rate in 2$\sigma$ regions: $0.00300_{-0.00300-0.00300}^{+0.00064+0.00511}$ for IDE1 ($b_1=1, b_2=1$ in Eqs.(\ref{eq:per-deltax},\ref{eq:per-deltac},\ref{eq:per-thetax},\ref{eq:per-thetac})), $0.00362_{-0.00362-0.00362}^{+0.00088+0.00590}$ for IDE2 ($b_1=1, b_2=0$), $0.00321_{-0.00321-0.00321}^{+0.00063+0.00547}$ for IDE3 ($b_1=0, b_2=1$), $0.00312_{-0.00312-0.00312}^{+0.00070+0.00555}$ for IDE4 ($b_1=0, b_2=0$). In 2$\sigma$ region, there is no evidence of the interaction between the dark sectors from the analysis of the large scale structure measurements (RSD and WL). The interaction rate is close to zero, which tells us that the constraint results between the perturbed $H$ and unperturbed $H$ coupled model are compatible with each other in the rest frame of either dark matter or dark energy. It would be also very hard to distinguish the different interacting dark energy models.

\begingroup
\squeezetable
\begin{center}
\begin{table}
\begin{tabular}{cccccc}
\hline\hline Parameters & Priors & IDE1 & IDE2 & IDE3 & IDE4 \\ \hline
$\Omega_bh^2$&[0.005,0.1]&
$0.02232_{-0.00025-0.00047}^{+0.00024+0.00046}$&
$0.02229_{-0.00025-0.00048}^{+0.00025+0.00049}$&
$0.02234_{-0.00025-0.00046}^{+0.00024+0.00048}$&
$0.02230_{-0.00025-0.00048}^{+0.00025+0.00049}$
\\
$\Omega_ch^2$&[0.01,0.99]&
$0.1140_{-0.0017-0.0039}^{+0.0021+0.0036}$&
$0.1139_{-0.0018-0.0039}^{+0.0021+0.0038}$&
$0.1140_{-0.0018-0.0039}^{+0.0021+0.0038}$&
$0.1141_{-0.0017-0.0039}^{+0.0021+0.0038}$
\\
$100\theta_{MC}$&[0.5,10]&
$1.04157_{-0.00055-0.00108}^{+0.00057+0.00108}$&
$1.04150_{-0.00055-0.00107}^{+0.00055+0.00111}$&
$1.04154_{-0.00054-0.00114}^{+0.00055+0.00111}$&
$1.04150_{-0.00058-0.00114}^{+0.00058+0.00111}$
\\
$\tau$&[0.01,0.8]&
$0.087_{-0.014-0.024}^{+0.012+0.026}$&
$0.088_{-0.013-0.024}^{+0.011+0.025}$&
$0.088_{-0.014-0.025}^{+0.013+0.027}$&
$0.088_{-0.014-0.024}^{+0.012+0.026}$
\\
$\xi_x$&[0,1]&
$0.00300_{-0.00300-0.00300}^{+0.00064+0.00511}$&
$0.00362_{-0.00362-0.00362}^{+0.00088+0.00590}$&
$0.00321_{-0.00321-0.00321}^{+0.00063+0.00547}$&
$0.00312_{-0.00312-0.00312}^{+0.00070+0.00555}$
\\
$w_x$&[-1,0]&
$-0.976_{-0.024-0.024}^{+0.005+0.041}$&
$-0.975_{-0.025-0.025}^{+0.006+0.040}$&
$-0.976_{-0.024-0.024}^{+0.005+0.041}$&
$-0.974_{-0.026-0.026}^{+0.006+0.043}$
\\
$n_s$&[0.5,1.5]&
$0.9772_{-0.0058-0.0109}^{+0.0056+0.0111}$&
$0.9761_{-0.0056-0.0109}^{+0.0056+0.0109}$&
$0.9771_{-0.0054-0.0109}^{+0.0054+0.0109}$&
$0.9761_{-0.0056-0.0110}^{+0.0057+0.0112}$
\\
${\rm{ln}}(10^{10}A_s)$&[2.4,4]&
$3.081_{-0.027-0.046}^{+0.023+0.049}$&
$3.083_{-0.026-0.044}^{+0.022+0.049}$&
$3.083_{-0.024-0.048}^{+0.024+0.052}$&
$3.084_{-0.026-0.046}^{+0.023+0.050}$
\\
\hline
$\Omega_x$&$-$&
$0.7079_{-0.0098-0.0188}^{+0.0099+0.0181}$&
$0.7071_{-0.0100-0.0199}^{+0.0101+0.0190}$&
$0.7080_{-0.0097-0.0199}^{+0.0097+0.0188}$&
$0.7065_{-0.0098-0.0207}^{+0.0100+0.0195}$
\\
$\Omega_m$&$-$&
$0.2921_{-0.0099-0.0181}^{+0.0098+0.0188}$&
$0.2929_{-0.0100-0.0190}^{+0.0100+0.0199}$&
$0.2920_{-0.0097-0.0188}^{+0.0096+0.0199}$&
$0.2935_{-0.0100-0.0195}^{+0.0099+0.0207}$
\\
$\sigma_8$&$-$&
$0.805_{-0.012-0.024}^{+0.012+0.023}$&
$0.805_{-0.011-0.023}^{+0.011+0.023}$&
$0.804_{-0.012-0.028}^{+0.015+0.029}$&
$0.805_{-0.012-0.023}^{+0.012+0.023}$
\\
$H_0$&$-$&
$68.50_{-0.83-1.63}^{+0.83+1.58}$&
$68.38_{-0.87-1.73}^{+0.86+1.62}$&
$68.50_{-0.77-1.73}^{+0.90+1.56}$&
$68.36_{-0.84-1.87}^{+0.97+1.68}$
\\
${\rm{Age}}/{\rm{Gyr}}$&$-$&
$13.788_{-0.036-0.072}^{+0.036+0.070}$&
$13.794_{-0.037-0.074}^{+0.038+0.073}$&
$13.787_{-0.037-0.067}^{+0.038+0.070}$&
$13.793_{-0.038-0.075}^{+0.038+0.073}$
\\
\hline\hline
\end{tabular}
\caption{In contrast to the mean values with $1,2\sigma$ errors of the parameters for four coupled model, where CMB from Planck 2015, BAO, SNIa, RSD, and WL data sets have been used. IDE1 is the coupled model with perturbed expansion rate $(b_2=1)$ when $Q^{\mu}\parallel u^{\mu}_{(c)}$ $(b_1=1)$; IDE2 is the coupled model without perturbed expansion rate $(b_2=0)$ when $Q^{\mu}\parallel u^{\mu}_{(c)}$ $(b_1=1)$; IDE3 is the coupled model with perturbed expansion rate $(b_2=1)$ when $Q^{\mu}\parallel u^{\mu}_{(x)}$ $(b_1=0)$; IDE4 is the coupled model without perturbed expansion rate $(b_2=0)$ when $Q^{\mu}\parallel u^{\mu}_{(x)}$ $(b_1=0)$.}
\label{tab:results-mean}
\end{table}
\end{center}
\endgroup

\begin{figure}
\centering
\subfigure{
\begin{minipage}[b]{0.44\textwidth}
\includegraphics[width=1\textwidth]{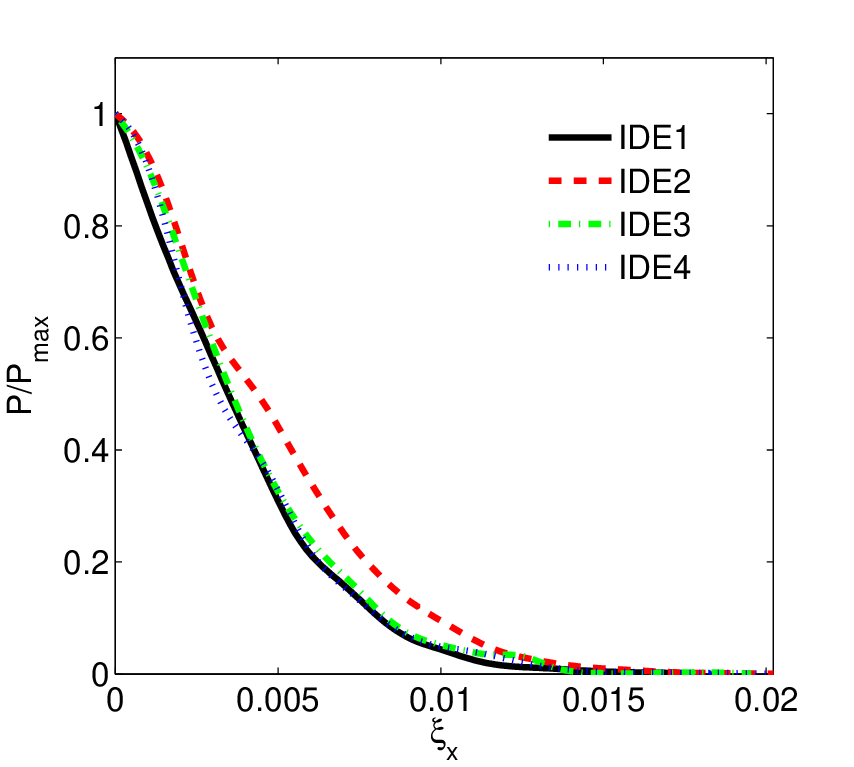} \\ 
\includegraphics[width=1\textwidth]{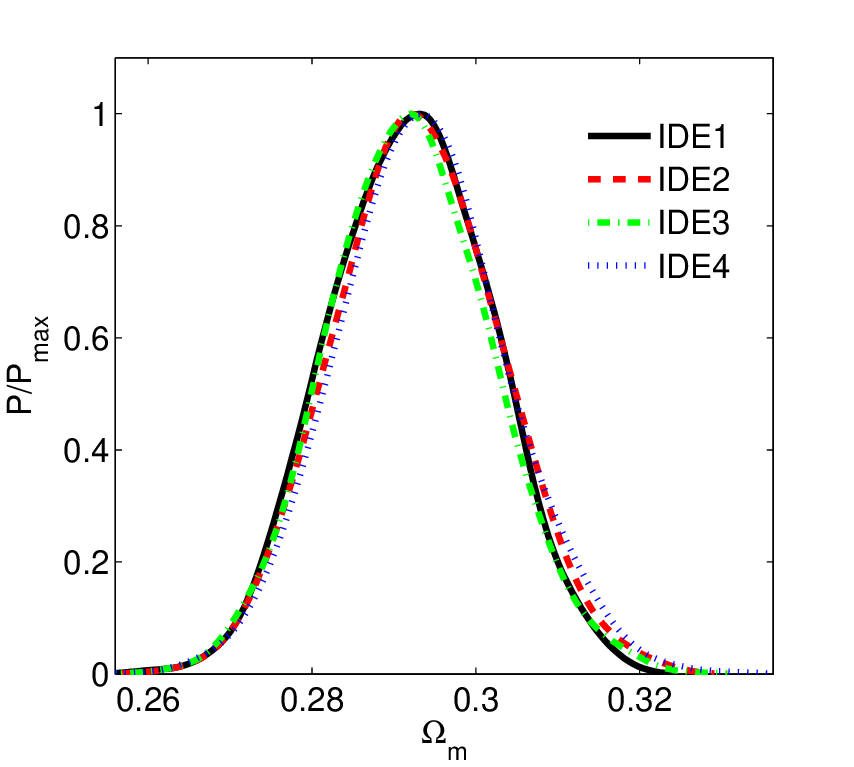}    
\end{minipage}
}
\subfigure{
\begin{minipage}[b]{0.44\textwidth}
\includegraphics[width=1\textwidth]{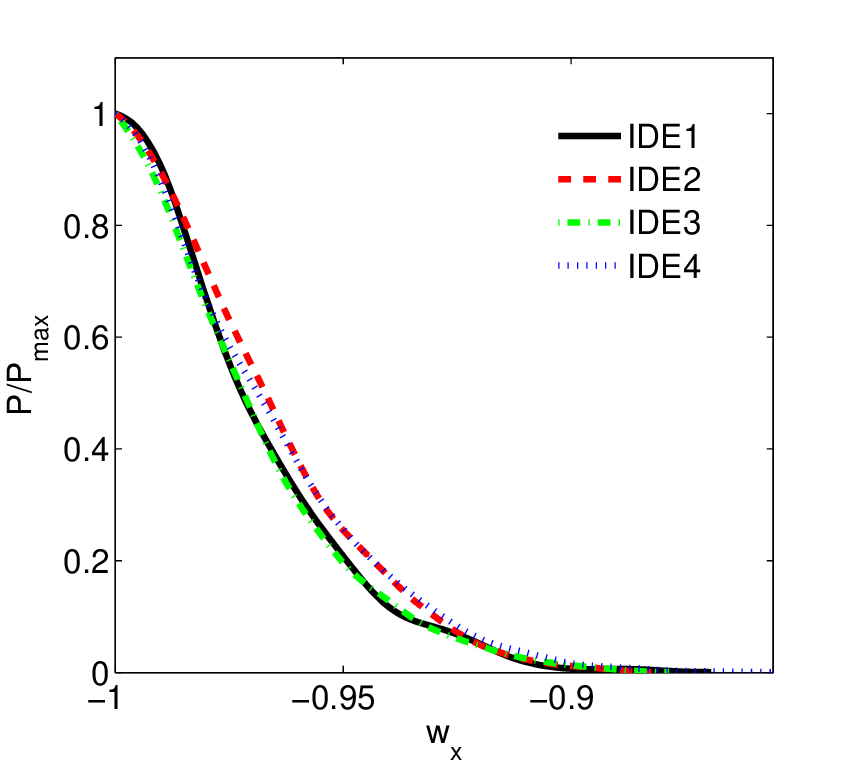} \\ 
\includegraphics[width=1\textwidth]{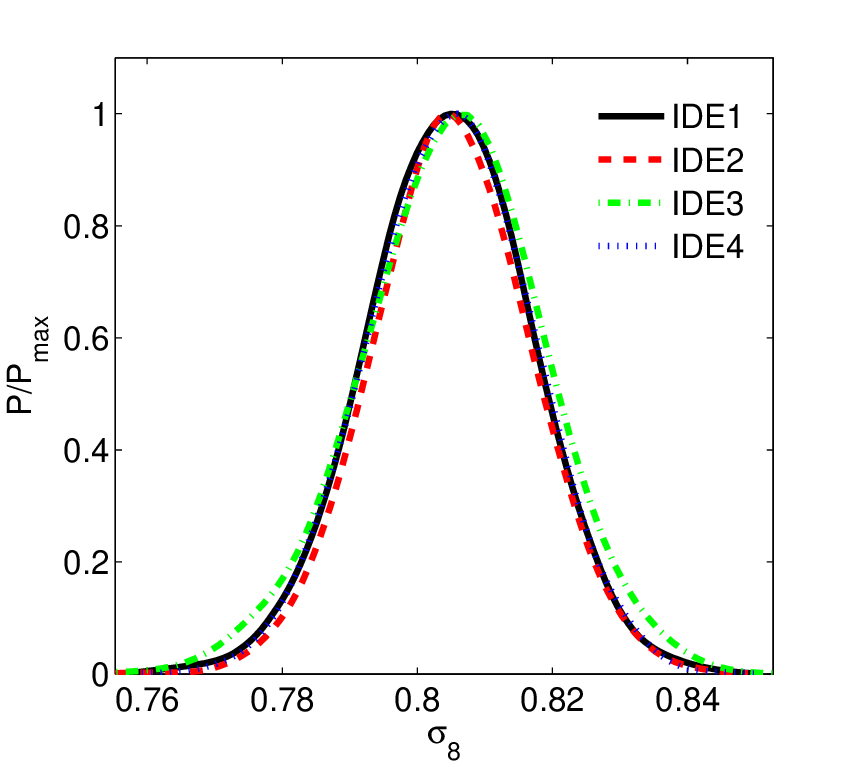}    
\end{minipage}
}
\caption{In contrast to the 1D marginalized distributions on the parameters $\xi_x$, $w_x$, $\Omega_m$, $\sigma_8$ of the four coupled models with 68\% C.L., 95 \% C.L., and 99.7\% C.L. between each other, where CMB from Planck 2015, BAO, SNIa, RSD, and WL data sets have been used. The line style of IDE1, IDE2, IDE3, IDE4 are, respectively, black solid, red dashed, green dotted-dashed, blue dotted.}
\label{fig:vs}
\end{figure}

\section{SUMMARY}

In this paper, we presented a numerical analysis of several phenomenological interacting dark energy models, where dark energy was treated as a fluid with a constant equation of state parameter. We considered that the energy transfer rate was proportional to the Hubble parameter and energy density of dark energy, $\bar{Q}=3H\xi_x\bar{\rho}_x$, the parameter $\xi_x$ was the interaction rate which characterized the coupled model and represented the coupling strength between dark matter and dark energy. For the Hubble expansion rate, $H$ denoted the total expansion rate (background plus perturbations). For the momentum transfer, we made the simplest physical choice, that was, there was no momentum transfer in the rest frame of either dark matter or dark energy. We have deduced the perturbation equations of dark sectors in the rest frame of dark matter or dark energy. Then, based on CMB from Planck 2015, BAO, SNIa, RSD, WL measurements, we conducted a full likelihood analysis for the four different coupled models. The jointing constraint results showed the interaction rate $\xi_x$ in 2$\sigma$ regions: $0.00300_{-0.00300-0.00300}^{+0.00064+0.00511}$ for IDE1, $0.00362_{-0.00362-0.00362}^{+0.00088+0.00590}$ for IDE2, $0.00321_{-0.00321-0.00321}^{+0.00063+0.00547}$ for IDE3, $0.00312_{-0.00312-0.00312}^{+0.00070+0.00555}$ for IDE4. There is no evidence of the interaction at 2$\sigma$ level from the analysis of the currently available cosmic observations. The interaction rate was close to zero, which told us that it would be very hard even in the future to distinguish different coupled models.

We only focused on the effects that can be calculated within linear perturbation theory. The use of matter power spectrum up to the smallest scales (largest wave numbers $k$) probed by the current large scale structure observations would require detailed modeling of the non-linear effects that may differ from the non-interacting models, as shown in Refs. \cite{ref:nonlinear1,ref:nonlinear2,ref:nonlinear3,ref:nonlinear4}. Yet there would remain ambiguity of the assumptions made on the bias function for the growth rate of matter, i.e., how the dark matter traces the visible matter on the smallest observable scales. In the near future, we would try to make some research work about interacting dark energy models on the non-linear scale.

\acknowledgements{This work is supported by National Natural Science Foundation of China under the Grants No. 11175077 and 11575075.}

\end{document}